\documentclass[twocolumn]{jpsj2}

\def\nn{\nonumber}
\newcommand{\bsigma}{\mbox{\boldmath $\sigma$}}

\title{Gauge field for edge state in graphene}

\author{
Ken-ichi \textsc{Sasaki}$^{1}$
\thanks{E-mail: sasaken@imr.tohoku.ac.jp},
Shuichi \textsc{Murakami}$^{2}$
\thanks{E-mail: murakami@appi.t.u-tokyo.ac.jp} and
Riichiro \textsc{Saito}$^{3}$
\thanks{E-mail: rsaito@flex.phys.tohoku.ac.jp}
}

\inst{
$^{1}$ Institute for Materials Research, Tohoku University, 
Sendai 980-8577, Japan \\
$^{2}$ Department of Applied Physics, University of Tokyo,
Hongo, Bunkyo-ku, \\
Tokyo 113-8656, Japan\\
$^{3}$ Department of Physics, Tohoku University and CREST, JST,\\
Sendai 980-8578, Japan
}

\recdate{\today}

\abst{ 
 By considering the continuous model for graphene, we analytically
 study a special gauge field for the edge state. 
 The gauge field explains the properties of the edge state 
 such as the existence only on the zigzag edge, 
 the partial appearance in the $k$-space, 
 and the energy position around the Fermi energy.
 It is demonstrated utilizing the gauge field that 
 the edge state is robust for surface reconstruction, and 
 the next nearest-neighbor interaction which breaks the particle-hole 
 symmetry stabilizes the edge state.
}

\kword{graphene, edge state, Weyl equation, gauge field, next
nearest-neighbor hopping, particle-hole symmetry}

\begin{document}

\maketitle

\section{Introduction}

A single layer of graphite or graphene is an element 
of various carbon-based materials 
such as carbon nanotubes.~\cite{SDD}
These materials are eminently suitable for future nanotechnology
and thus fundamental understanding of the electronic properties 
is indispensable.
The electronic property of graphene is mainly determined 
by the $\pi$-electrons around the Fermi energy 
at the hexagonal corners (the K and K$'$ points) 
of the two-dimensional Brillouin zone.
It is known that the eigenstates are classified into
extended states and localized states.
The edge state existing near the zigzag edge is an example
of the localized state.~\cite{Fujita}
The tight-binding (TB) lattice model has shown that 
the edge state exists near the zigzag edge 
while no localized state exists near the armchair
edge.~\cite{Fujita,Nakada}
Recently, 
by scanning tunneling microscopy and spectroscopy (STM/STS) experiments,
Niimi {\it et al.}~\cite{Niimi} and 
Kobayashi {\it et al.}~\cite{Kobayashi}, respectively, 
observed the edge state near the zigzag edge 
and the energy position.

The edge shape determines 
the boundary condition for the wave function.
It has been shown in the studies of single-walled carbon nanotubes 
(SWNTs) that 
a boundary condition can be changed 
by an external magnetic field and uniform lattice deformations.
Using the continuous model of graphene,~\cite{SW} 
Ajiki and Ando~\cite{AA} showed that 
a uniform magnetic field parallel to a SWNT axis 
affects the energy band structure of extended states.
The magnetic field is expressed by a gauge field 
which changes the boundary condition around a tube axis through the
Aharonov-Bohm (AB) effect.
Zaric {\it et al.}~\cite{Zaric} observed optical signatures of the
AB effect.
Kane and Mele~\cite{KM} introduced another homogeneous gauge field 
in the continuous model.
The gauge field represents uniform lattice deformations 
such as uniform bend, twist and curvature.
The gauge field changes the boundary condition around a tube axis
through a generalized AB effect and 
makes a mini gap in chiral {\it metallic} SWNTs.
The curvature-induced mini gap was observed by STS experiment by 
Ouyang {\it et al}.~\cite{Ouyang}
The perturbations above modify the boundary condition for extended
states and are similar in that they are represented by gauge
fields in the continuous model.
Is it possible to change the boundary condition for the edge state? 
To answer this question, a generalization of the gauge fields to
local fields is necessary since the edge states appear locally 
around the edge.
In the previous paper,~\cite{SKS}
we generalize the gauge field introduced by Kane and Mele,
to include a local lattice deformation of graphene.
The deformation-induced gauge field 
accounts for the local modulation of the energy band gap,~\cite{SKS}
which was observed in some pea-pod-like structures 
by Lee {\it et al}.~\cite{Lee}
In this paper, using the continuous model, 
we show that the edge state can be formed 
by a deformation-induced ``magnetic'' field and 
we clarify why the edge state is formed in the zigzag edge, but 
not in the armchair edge.
We explain various properties of the edge states, and
examine how the edge state are affected by perturbations.

Here we briefly introduce the continuous model.
In an undeformed graphene,
the Hamiltonian for $\pi$-electrons around the Fermi point 
are given by two-component Weyl equation.~\cite{SW}
Two components of the wave function represent 
the two sublattices in the unit cell.
The Weyl equation gives two-dimensional linear $k$ energy dispersion
relation around the K and K$'$ points in the Brillouin zone. 
The Weyl equation allows us to treat a variety of perturbations 
in a unified way for studying transport properties,
scattering process around impurities, 
lattice defects and topological defects.~\cite{Ando}

This paper is organized as follows.
In \S~\ref{sec:cont-theo}, 
we define a gauge field for a locally deformed graphene system.
Then we solve the continuous model (Weyl equation with the gauge field) 
to obtain the edge state.
In \S~\ref{sec:app}, 
we present several applications of the continuous model.
The effects of 
a lattice deformation around the edge, 
the next nearest-neighbor interaction, 
and an external magnetic field 
on the edge state are examined.
In \S~\ref{sec:discuss}, 
we discuss and summarize the results.
We derive the continuous model in Appendix~\ref{app:gauge}.
In Appendix~\ref{app:valid}, 
we analyze a locally deformed graphene system using the TB model.

\section{Continuous Model}\label{sec:cont-theo}

\begin{figure}[htbp]
 \begin{center}
  \includegraphics[scale=0.4]{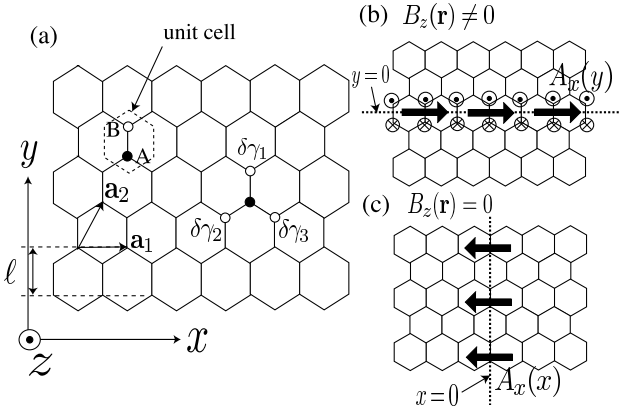}
 \end{center}
 \caption{(a) Lattice structure of graphene.
 The unit cell consists of A (closed circle) and B (open circle)
 sublattices, which leads to the pseudo-spin structure of the wave
 function.
 Local modulations of the hopping integral is defined by 
 $\delta \gamma_a(\mathbf{r})$ ($a=1,2,3$).
 Two examples of local deformation (borders) and resultant
 deformation-induced gauge fields are shown in (b) and (c).
 In (b), the gauge field (arrows) gives a finite deformation-induced
 magnetic field (flux) illustrated by $\odot$ and $\otimes$ while no
 deformation-induced magnetic field is present in (c).} 
 \label{fig:graphene}
\end{figure}

We consider an undeformed graphene 
as shown in Fig.~\ref{fig:graphene}(a).
The unit cell consists of 
A sublattice ($\bullet$) and B sublattice ($\circ$),
and each sublattice is spanned by primitive vectors 
$\mathbf{a}_1$ and $\mathbf{a}_2$.
$|2\mathbf{a}_2-\mathbf{a}_1| \equiv 2 \ell$
is the unit length along the $y$-axis.
A $\pi$-electron hops from one site to the nearest-neighbor sites 
with the hopping integral $-\gamma_0$ ($\approx -3$ eV)
in each hopping process.
Let $v_F (\equiv \gamma_0 \ell/\hbar)$ be the Fermi velocity and 
$\hat{\mathbf{p}} \equiv (\hat{p}_x,\hat{p}_y)$ the momentum operator.
The Hamiltonian 
for $\pi$-electrons around the K point is then written as~\cite{SW}
\begin{align}
 {\cal H}_{\rm K}^0 = v_F
 \bsigma' \cdot \hat{\bf{p}} = v_F
 \begin{pmatrix}
  0 & -\hat{p}_x -i \hat{p}_y \cr
  -\hat{p}_x + i \hat{p}_y & 0 
 \end{pmatrix},
 \label{eq:weyl0}
\end{align}
where $\bsigma' \equiv (-\sigma_x,\sigma_y)$ and 
$\sigma_i$ ($i=x,y,z$) are the Pauli matrices.
The Hamiltonian operates on a two-component wave function
$\psi^{\rm K}({\mathbf r}) = 
{}^t(\psi^{\rm K}_{\rm A}({\bf r}),\psi^{\rm K}_{\rm B}({\bf r}))$.
This two-component character is referred to as the {\it pseudo-spin};
hence, $\psi_{\rm A}^{\rm K}({\bf r})$ and $\psi^{\rm K}_{\rm B}({\bf r})$
are the pseudo-spin up and down states, respectively.
The energy eigenequation 
${\cal H}_{\rm K}^{0}\psi_E^{\rm K}({\mathbf r})=E \psi_E^{\rm K}({\mathbf r})$
is the Weyl equation.

When the lattice is deformed locally, 
the nearest-neighbor hopping integral depends on the position as
$-\gamma_0 + \delta \gamma_a(\mathbf{r})$ ($a=1,2,3$) where 
$a$ denotes three nearest-neighbor B sites from an A atom 
(see Fig.~\ref{fig:graphene}(a)).
The Hamiltonian can be expressed by
\begin{align}
 {\cal H}_{\rm K} = v_F
 \bsigma' \cdot (\hat{\bf{p}}-\bf{A}({\bf r})),
 \label{eq:weyl}
\end{align}
where the vector field
$\mathbf{A}(\mathbf{r})=(A_x(\mathbf{r}),A_y(\mathbf{r}))$ 
is the deformation-induced gauge field.
For a weak lattice deformation 
satisfying $|\delta \gamma_a(\mathbf{r})| \ll \gamma_0$, 
the gauge field is expressed by 
a linear combination of $\delta \gamma_a(\mathbf{r})$ as~\cite{SKS}
(see Appendix~\ref{app:gauge})
\begin{align}
 \begin{split}
  & v_F A_x({\bf r}) = \delta \gamma_1({\bf r}) 
  - \frac{1}{2} \left(
  \delta \gamma_2({\bf r})+ \delta \gamma_3({\bf r}) \right), \\
  & v_F A_y({\bf r}) = \frac{\sqrt{3}}{2} 
  \left( \delta \gamma_2({\bf r})- \delta \gamma_3({\bf r}) \right).
 \end{split}
 \label{eq:gauge}
\end{align}

Here, we give two examples of local lattice deformation on a line
(Figs.~\ref{fig:graphene}(b) and~\ref{fig:graphene}(c)).
From now we call this type of lattice deformation {\it border}.
The example of Fig.~\ref{fig:graphene}(b)
is a modification of C-C bond lengths in a row at $y=0$,
namely, $\delta \gamma_1(\mathbf{r})\ne 0$ at $y=0$ and 
$\delta \gamma_2(\mathbf{r}) = \delta \gamma_3(\mathbf{r}) = 0$.
From eq.(\ref{eq:gauge}), 
we obtain $v_F \mathbf{A}(y) = (\delta \gamma_1(y),0)$.
The example of Fig.~\ref{fig:graphene}(c) is that
$\delta \gamma_1(\mathbf{r})=0$ and 
$\delta \gamma_2(\mathbf{r}) = \delta \gamma_3(\mathbf{r}) \ne 0$ 
at $x=0$.
We obtain $v_F \mathbf{A}(x) = (-\delta \gamma_2(x),0)$ for this case. 
Figures~\ref{fig:graphene}(b) and~\ref{fig:graphene}(c)
correspond to the zigzag edge and the armchair edge if we take
$\delta \gamma_1= \gamma_0$ and 
$\delta \gamma_2=\delta \gamma_3= \gamma_0$,
respectively.
The difference between the two gauge fields 
becomes clear 
by defining the deformation-induced ``magnetic'' field, 
\begin{align}
 B_z(\mathbf{r}) \equiv
 \frac{\partial A_y(\mathbf{r})}{\partial x} 
 - \frac{\partial A_x(\mathbf{r})}{\partial y},
\end{align}
which is perpendicular to the graphene plane
${\bf B}(\mathbf{r})= (0,0,B_z(\mathbf{r}))$.
It should be noted that it is not a real magnetic field.
It is called a magnetic field only because it enters the Hamiltonian 
in the similar way as the real magnetic field.
We have a finite deformation-induced magnetic field 
for Fig.~\ref{fig:graphene}(b), 
and zero magnetic field for Fig.~\ref{fig:graphene}(c).
The gauge field which gives zero magnetic field can be removed from
eq.(\ref{eq:weyl}) by a gauge transformation.~\cite{SKS}
As shown in Fig.~\ref{fig:graphene}(b), 
the magnetic field appears locally around $y=0$;
$B_z(y)$ is negative at $y < 0$ as illustrated by $\otimes$ 
and is positive at $y > 0$ as $\odot$.
The deformation-induced magnetic field will account for 
the presence of edge states not at the armchair edge 
but at the zigzag edge.
The importance of the magnetic field can be seen by the fact that the
energy eigenstate of eq.(\ref{eq:weyl}) also satisfies
${\cal H}_{\rm K}^2 \psi^{\rm K}_E(\mathbf{r}) = E^2 \psi^{\rm
K}_E(\mathbf{r})$:
\begin{align}
 v_F^2 \left\{ (\hat{\mathbf{p}}-\mathbf{A}(\mathbf{r}))^2 + \hbar
 B_z(\mathbf{r}) \sigma_z \right\} 
 \psi^{\rm K}_E(\mathbf{r})
 = E^2 \psi^{\rm K}_E(\mathbf{r}).
 \label{eq:second-H}
\end{align}
${\cal H}_{\rm K}^2$ has only diagonal-components 
in the 2$\times$2 matrix.
One can see that $B_z(\mathbf{r})$ in eq.(\ref{eq:second-H})
works as a potential and affects the electronic properties.
This is because the diagonal element is similar to 
the non-relativistic Hamiltonian
${\cal H}_{\rm nr}=(1/2m)(\hat{\bf{p}}-\mathbf{A}(\mathbf{r}))^2 \pm
V(\mathbf{r})$ 
where the sign in front of $V(\mathbf{r})$ 
corresponds to the pseudo-spin and $m$ is a mass.

\begin{figure}[htbp]
 \begin{center}
  \includegraphics[scale=0.45]{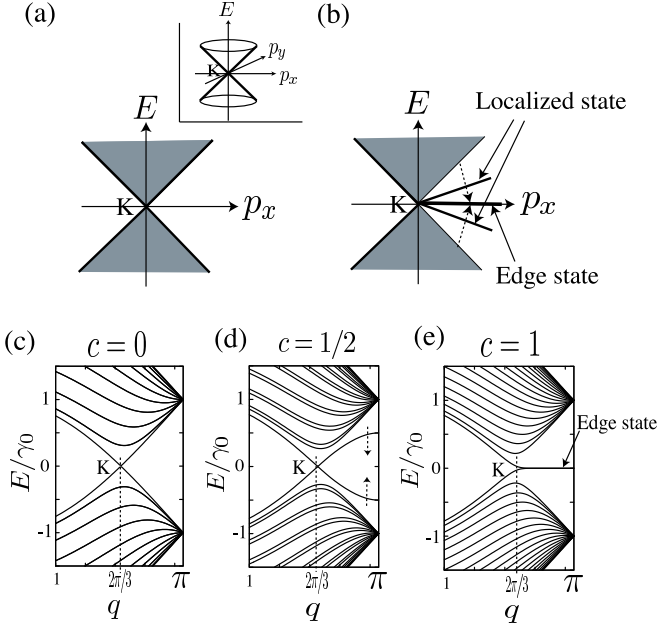}
 \end{center}
 \caption{(a) Band structure $(p_x,E)$ around the K point of the
 Weyl equation without the deformation-induced gauge field and (b) that
 with the gauge field at the border.
 The shaded regions in (a) and (b) represent 
 the spectrum of extended states. 
 The (two-dimensional) dispersion relation around the K point 
 for graphene without the
 border gives two cones (representing the linear $k$ dispersion) whose
 apex is the K point as is shown in the inset of (a).
 The band structure of (a) is a one-dimensional projection (onto $p_y$
 plane) of the linear $k$ dispersion.
 In (b), the energy dispersion for the localized state is represented by
 solid lines.
 When the bonds at $y=0$ become disconnected and zigzag edges appear,
 the energy eigenvalues of the localized state converge to zero ($E\to 0$)
 for any $p_x(>0)$ and forms a flat energy band of edge state.
 Using the TB model, we plot the band structure for an undeformed
 graphene in (c), for a zigzag ribbon in (e), and for a graphene with the
 border (with weakened hopping for the C-C bonds at $y=0$) in (d).
 $q$ in (c), (d) and (e) is the momentum parallel to the border and
 $q=2\pi/3$ is the K point.
 }
 \label{fig:cone}
\end{figure}

We derive the edge states from the Weyl equation with the 
gauge field of Fig.~\ref{fig:graphene}(b), 
corresponding to the zigzag edge.
Before going into detail, we outline the results here.
We will show that there are localized states in the energy spectrum
and the energy dispersion appears as 
the two solid lines at $p_x > 0$ shown in Fig.~\ref{fig:cone}(b).
The velocity becomes small with increasing the gauge field 
and it becomes zero when the gauge field is sufficiently strong.
The result of the continuous model 
reproduces the flat band of the edge states 
shown in Fig.~\ref{fig:cone}(e) which is calculated 
from the TB model~\cite{Fujita} 
with the zigzag edge. 

We assume that $\mathbf{A}(\mathbf{r})$ of Fig.~\ref{fig:graphene}(b) 
is quite localized within $|y|<\xi_g$ 
where $\xi_g$ is a length of the order of lattice spacing, namely, 
$\mathbf{A}(\mathbf{r})=(A_x(y),0)$ and 
$A_x(y)=0$ for $|y| \ge \xi_g$ in eq.(\ref{eq:weyl}).
We parameterize the localized energy eigenstate as
\begin{align}
 \psi^{\rm K}_E(\mathbf{r}) = 
 N' \exp(i\frac{p_x x}{\hbar}) e^{-G(y)}
 \begin{pmatrix}
  e^{+g(y)} \cr e^{-g(y)}
 \end{pmatrix},
 \label{eq:wfunc}
\end{align}
where $N'$ is a normalization constant.
The pseudo-spin modulation is 
represented by $g(y)$, and 
the wave vector $p_x$ is a good quantum number 
because of the translational symmetry along the border. 
Putting eq.(\ref{eq:wfunc}) to the energy eigenequation,
${\cal H}_{\rm K}\psi^{\rm K}_E(\mathbf{r})=E \psi^{\rm
K}_E(\mathbf{r})$, we obtain
\begin{align}
 \begin{split}
  & p_x - A_x(y) -\hbar \frac{d}{dy} \left(G(y)+g(y)\right) =
  - \frac{E}{v_F} e^{+2g(y)}, \\ 
  & p_x - A_x(y) +\hbar \frac{d}{dy} \left(G(y)-g(y)) \right) = 
  - \frac{E}{v_F} e^{-2g(y)}.
 \end{split}
 \label{eq:w+g-d}
\end{align}
By summing and subtracting the both sides of eq.(\ref{eq:w+g-d}), 
we rewrite the energy eigenequation as 
\begin{align}
 \begin{split}
  & p_x -A_x(y) - \hbar \frac{dg(y)}{dy} 
  = - \frac{E}{v_F} \cosh(2g(y)), \\
  & \hbar \frac{dG(y)}{dy} = \frac{E}{v_F} \sinh(2g(y)).
 \end{split} 
 \label{eq:weyl+g}
\end{align}

\begin{figure}[htbp]
 \begin{center}
  \includegraphics[scale=0.45]{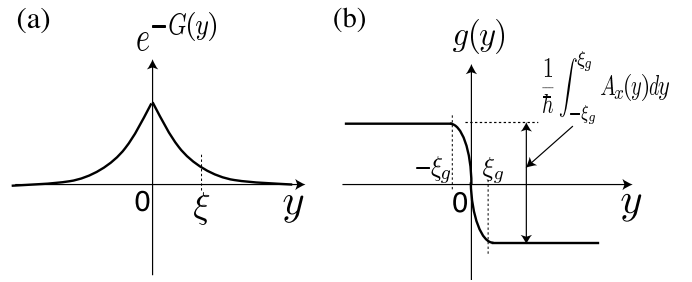}
 \end{center}
 \caption{(a) The amplitude of the wave function, $\exp(-G(y))$, of a 
 localized state whose localization length is $\xi$.
 (b) The pseudo-spin modulation part, $g(y)$.
 From eq.(\ref{eq:const-g(y)}), $g(y)$ is a constant 
 for $|y| \ge \xi_g$, and abruptly changes across the border ($y=0$).  
 }
 \label{fig:Gandg}
\end{figure}

Since we are considering a localized solution around the border,
we assume $G(y) \sim |y|/\xi$ 
where $\xi$ $(>0)$ is the localization length.
When $G(y) \sim |y|/\xi$, 
the solution of the second equation of eq.(\ref{eq:weyl+g}) is given by 
\begin{align}
 g(y) = \left\{
 \begin{array}{@{\,}ll}
  -\frac{1}{2} \sinh^{-1} \left( \frac{\hbar v_F}{\xi E}
  \right) & (y \le -\xi_g), \\
  +\frac{1}{2} \sinh^{-1} \left( \frac{\hbar v_F}{\xi E}
  \right) & (y \ge \xi_g). 
 \end{array} \right.
 \label{eq:const-g(y)}
\end{align}
The functions $G(y)$ and $g(y)$ 
are schematically shown in 
Fig.~\ref{fig:Gandg}(a) and~\ref{fig:Gandg}(b), respectively.
The sign of $g(y)$ changes across the border; 
this sign change means that the pseudo-spin flips at the border. 
The flip is induced by the gauge field $A_x(y)$.
To see this, we integrate the first equation of eq.(\ref{eq:weyl+g})
from $y=-\xi_g$ to $\xi_g$, and acquire
\begin{align}
 - \int_{-\xi_g}^{\xi_g} \frac{dg(y)}{dy} dy
 = \frac{1}{\hbar} \int_{-\xi_g}^{\xi_g} A_x(y) dy.
 \label{eq:w-s-integ}
\end{align}
We have neglected other terms, since they are proportional to
$\xi_g$ and become zero in the limit of $\xi_g =0$.
By putting eq.(\ref{eq:const-g(y)}) to 
eq.(\ref{eq:w-s-integ}), we find
\begin{align}
 -\sinh^{-1} \left( \frac{\hbar v_F}{\xi E} \right) =
 \frac{1}{\hbar} \int_{-\xi_g}^{\xi_g} A_x(y) dy.
 \label{eq:g-ene}
\end{align}
When the right-hand side of eq.(\ref{eq:g-ene}) is large,
one obtain from eq.(\ref{eq:const-g(y)}) that 
$g(y) \gg 0$ for $y \le -\xi_g$ and $g(y) \ll 0$ for $y \ge \xi_g$.
In this case, 
the localized state is approximately a pseudo-spin-up state
$\psi^{\rm K}_{E}(\mathbf{r}) \propto {}^t (1,0)$
for $y \le -\xi_g$ and a pseudo-spin-down state
$\psi^{\rm K}_{E}(\mathbf{r}) \propto {}^t (0,1)$ for $y \ge \xi_g$.
Hence, a strong gauge field  at the border makes  
pseudo-spin-polarized localized states.
Since the polarization of the pseudo-spin means that 
the wave function has amplitude only A (or B) sublattice,
this result agrees with the result by the TB model for the
edge state.~\cite{Fujita}

Having described the wave function of the localized state, 
let us now calculate $E$ and $\xi$.
To this end, we use the first equation of eq.(\ref{eq:weyl+g}) 
for $|y| \ge \xi_g$
and obtain 
\begin{align}
 \frac{E}{v_F}
 = \frac{-p_x}{\cosh \left( \displaystyle 
 \frac{1}{\hbar} \int_{-\xi_g}^{\xi_g} A_x(y) dy \right)}.
 \label{eq:ene}
\end{align}
Moreover, using eq.(\ref{eq:g-ene}), we find
\begin{align}
 \frac{\hbar}{\xi}
 = p_x \tanh \left( \displaystyle 
 \frac{1}{\hbar} \int_{-\xi_g}^{\xi_g} A_x(y) dy \right).  
 \label{eq:xi}
\end{align}
In addition to this localized state, 
there is another localized state for the same $p_x$ with 
the same $\xi$ but with an opposite sign of $E$.
This results from a particle-hole symmetry of the Hamiltonian;
$\sigma_z {\cal H}_{\rm K} \sigma_z 
=-{\cal H}_{\rm K}$.
By the particle-hole symmetry operation, 
the wave function is transformed as 
$\psi^{\rm K}_{-E}(\mathbf{r}) = \sigma_z \psi^{\rm K}_E(\mathbf{r})$.

The normalization condition of the wave function requires that 
$\xi$ should be positive, which 
restricts the value of $p_x$ in eq.(\ref{eq:xi}).
Indeed, when $A_x(y)$ is positive, 
eq.(\ref{eq:xi}) means that the localized states appear only 
at $p_x>0$ around the K point. 
This is the reason why 
the localized states appear in the energy spectrum
only in one side around the K point in Fig.~\ref{fig:cone}(b).
On the other hand,
the Hamiltonian around the K$'$ point is expressed by
\begin{align}
 {\cal H}_{\rm K'}=
 v_F \bsigma \cdot (\hat{\mathbf{p}} + \mathbf{A}(\mathbf{r})),
 \label{eq:weyl-K'}
\end{align}
where $\bsigma \equiv (\sigma_x,\sigma_y)$.
The different signs in front of $\mathbf{A}(\mathbf{r})$
in eqs.(\ref{eq:weyl}) and (\ref{eq:weyl-K'})
guarantee the time-reversal symmetry.
Because of the different signs,
a similar argument as we used for the K point concludes that 
the localized state appears $p_x < 0$ around the K$'$ point.
Furthermore, 
when $(1/\hbar) \int_{-\xi_g}^{\xi_g} A_x(y) dy \gg 0$, 
$E$ in eq.(\ref{eq:ene}) becomes zero.
The zero energy eigenvalue between the K and K$'$ points in the band
structure corresponds to the flat energy band of the edge
state.~\cite{Fujita}

When $A_x(y)$ is negative ($A_x(y)\ll 0$),  
a flat energy band appears in the opposite side: $p_x < 0$
around the K point and $p_x > 0$ around the K$'$ point.
This condition, $A_x(y) \ll 0$, 
corresponds to the Klein's edges~\cite{Klein}
which are obtained by removing A or B sites 
out of the zigzag edges having A or B sites.
Calculations on the TB model with the Klein's edges~\cite{Klein} also
agree with these results obtained here.

There are also extended states in addition to the edge states.
The energy dispersion relation of extended states 
can be obtained as $(E/v_F)^2 = p_x^2 + p_y^2$, 
by setting $G(y) \sim -ip_y y/\hbar $ ($|y| \ge \xi_g$) in
eqs.(\ref{eq:wfunc}) and (\ref{eq:weyl+g}) 
where $p_y$ is a real number.
The calculated energy bands are given by $|E|>v_F|p_x|$, 
shown as a shaded region in Fig.~\ref{fig:cone}(b). 
It also agrees with the TB calculation shown in Fig.~\ref{fig:cone}(e).
One can then regard the localized state as a state with a
complex wavenumber as follows.
Since $1-\tanh^2 x=1/\cosh^2 x$, 
we see that 
the energy dispersion relation of the localized state 
between $E$ and $\xi$ becomes
\begin{align}
 \left( \frac{E}{v_F} \right)^2 = p_x^2 - \left( \frac{\hbar}{\xi}
 \right)^2, \ \ (|y| \ge \xi_g),
 \label{eq:energydispersion}
\end{align}
which is the same as the linear dispersion relation 
$(E/v_F)^2 = p_x^2 + p_y^2$ 
if one replaces $p_y$ with $i(\hbar/\xi)$.

We have shown that
three basic properties of the edge states, 
i.e., the pseudo-spin polarization,
the dependence on the momentum,
and the flat energy band,
obtained previously by the TB model,~\cite{Fujita} 
can be explained in terms of the gauge field. 
In order to quantitatively compare 
the present theory with the TB model,
we have performed a TB calculation for the geometry of
Fig.~\ref{fig:graphene}(b) with changing $\delta \gamma_1$. 
Here, we introduce an adiabatic parameter $c$ by 
$\delta \gamma_1 = c \gamma_0$, that is, 
$c=0$ and $c=1$ correspond to no deformation and the zigzag edge,
respectively. 
In Figs.~\ref{fig:cone}(c),~\ref{fig:cone}(d) and~\ref{fig:cone}(e), 
we plot the band structure for $c=0$, $c=1/2$, and $c=1$.
Comparing these figures with the results of the continuous model, one
can find a good correspondence 
between the TB model and continuous model. 
Moreover, we analytically find (see Appendix~\ref{app:valid})
\begin{align}
 \begin{split}
  & \frac{|E|}{v_F} = \frac{|p_x|}{\cosh (-\ln(1-c))} + 
  {\cal O}(l p_x^2/\hbar), \\
  & \frac{\hbar}{\xi} = p_x \tanh (- \ln (1-c)) + 
  {\cal O}(l p_x^2/\hbar),
 \end{split}
 \label{eq:lat-res-fin}
\end{align}
for localized states around the K point.
Thus, by comparing eq.(\ref{eq:lat-res-fin}) with 
eqs.(\ref{eq:ene}) and (\ref{eq:xi}),
we conclude that the TB model and the continuous model 
agree with each other near the K point ($p_x l/\hbar \ll 1$),
by the following relationship
\begin{align}
 \frac{1}{\hbar} \int_{-\xi_g}^{\xi_g} A_x(y) dy 
 = - \ln (1-c).
 \label{eq:ans-gauge}
\end{align}
The right-hand side diverges when $c \to 1$, which 
reproduces the flat energy band ($E \to 0$) in eq.(\ref{eq:ene}) 
and gives $\xi/\hbar = p_x^{-1}$ in eq.(\ref{eq:xi}).

When a deformed graphene is located in an external magnetic field,
the Hamiltonians for the K and K$'$ points 
are given by replacing $\hat{\mathbf{p}}$
in eqs.(\ref{eq:weyl}) and (\ref{eq:weyl-K'})
with $\hat{\mathbf{p}} -e\mathbf{A}^{\rm em}(\mathbf{r})$ as
\begin{align}
 \begin{split}
  & {\cal H}_{\rm K} = v_F
  \bsigma' \cdot (\hat{\mathbf{p}}-e \mathbf{A}^{\rm em}(\mathbf{r})
  -\mathbf{A}(\mathbf{r})), \\
  & {\cal H}_{\rm K'} =
  v_F \bsigma \cdot (\hat{\mathbf{p}}-e \mathbf{A}^{\rm em}(\mathbf{r})
  + \mathbf{A}(\mathbf{r})),
 \end{split}
 \label{eq:weyl-KK'}
\end{align}
where $e$ denotes the electron charge and 
$\mathbf{A}^{\rm em}(\mathbf{r})$ is the
gauge field for the external magnetic field.
The $z$-component of the external magnetic field is given by
\begin{align}
 B_z^{\rm em}(\mathbf{r}) \equiv 
  \frac{\partial A_y^{\rm em}(\mathbf{r})}{\partial x}
  - \frac{\partial A_x^{\rm em}(\mathbf{r})}{\partial y}.
\end{align}
The same signs in front of $e \mathbf{A}^{\rm em}(\mathbf{r})$ 
in eq.(\ref{eq:weyl-KK'})
reflect the violation of time-reversal symmetry.
Due to the similarity between 
the external magnetic field and 
the deformation-induced magnetic field in eq.~(\ref{eq:weyl-KK'}),
one may expect that the edge states can be induced also by
$\mathbf{A}^{\rm em}(\mathbf{r})$.
However, as we will see in \S~\ref{sec:app}, 
the deformation-induced magnetic field for the edge state
corresponds approximately to one flux quantum in a hexagonal unit cell, 
and is thus extremely strong
($\sim 10^5$ T), compared with 
the external magnetic field.

\section{Application of Continuous Model}\label{sec:app}

In this section, we apply the continuous model
to the following three examples.

First, we consider an effect of surface reconstruction (SR) 
around the zigzag edge on the edge state.
Since the SR is an additional lattice deformation around the edge,
it can be expressed by an additional gauge field in the continuous model.
Let $\delta \mathbf{A}^{\rm SR}(\mathbf{r}) 
= (\delta A^{\rm SR}_x, \delta A^{\rm SR}_y)$ be the
gauge field for SR. The perturbation is then represented by
\begin{align}
 \delta {\cal H}^{\rm SR}(\mathbf{r}) = v_F 
 \begin{pmatrix}
  0 & \delta A^{\rm SR}_x + i \delta A^{\rm SR}_y \cr
  \delta A^{\rm SR}_x -i \delta A^{\rm SR}_y & 0
 \end{pmatrix}.
 \label{eq:perterb-a}
\end{align}
To calculate the energy shift 
by $\delta {\cal H}^{\rm SR}(\mathbf{r})$,
we recall that the edge state 
($\psi^{\rm K}_{E=0}$)
is a pseudo-spin polarized state.
Since $\delta {\cal H}^{\rm SR}(\mathbf{r})$ 
only mixes the wave function on A and B sublattices to each other,
it is shown within the first-order perturbation theory that 
$\delta {\cal H}^{\rm SR}(\mathbf{r})$ does not shift 
the energy of the edge state, namely, 
$\Delta E^{\rm SR} \equiv \langle \psi^{\rm K}_{0}|\delta {\cal
H}^{\rm SR}|\psi^{\rm K}_{0} \rangle \approx 0$.
This result is consistent with the result of 
Fujita {\it et al}.~\cite{Fujita2}, who found that 
additional lattice distortion has little effect on 
the energies of the edge states.

Second, we consider {\it next} nearest-neighbor (nnn) hopping.
The nnn interaction 
decreases the energies of (i.e., stabilizes) 
the edge states, as shown by the TB calculation.~\cite{SMS} 
It accounts for the STM/STS measurements~\cite{Niimi,Kobayashi},
in which the edge state appears about 20 meV 
{\it below} the Fermi energy.
We note that the importance of the nnn on a localized state around a 
defect is pointed out by Peres {\it et al.}~\cite{Peres}

The reason for the stabilization is that the nnn interaction mixes 
the wave function on the same sublattices, namely, 
it gives diagonal terms in eq.(\ref{eq:weyl}).
Here, we propose a simple way 
to include the nnn hopping process into the continuous model,
and then calculate the energy shift of the edge states.
Let  $-\gamma_n$ ($\approx -0.1 \gamma_0$) 
denote the nnn hopping integral.~\cite{Porezag}
For an undeformed graphene,
the nnn interaction can be included by adding 
$- \gamma_n (l/\hbar)^2 \hat{\bf{p}}^2$ to eq.(\ref{eq:weyl0})
as~\cite{SMS}
\begin{align}
 v_F \bsigma' \cdot \hat{\bf{p}} 
 - \gamma_n \left( \frac{\ell}{\hbar} \right)^2 \hat{\bf{p}}^2.
 \label{eq:w+n}
\end{align}
The border is incorporated by 
replacing $\hat{\bf{p}}$ 
with $\hat{\bf{p}}-\mathbf{A}(\mathbf{r})$ in eq.(\ref{eq:w+n}).
As a result, we can write the total Hamiltonian as
\begin{align}
 \mathbf{H}_{\rm K} = {\cal H}_{\rm K}
 -\gamma_n \left( \frac{\ell}{\hbar} \right)^2
 (\hat{\bf{p}}-\mathbf{A}(\mathbf{r}))^2.
 \label{eq:w+n2}
\end{align}
It is noted that 
this replacement is not always correct in general
because the modulation of the nnn hopping integral is generally
independent of the nearest neighbor one.
In the present case, however, we adopt this replacement as a 
simplest approximation.
Using 
$({\cal H}_{\rm K}/v_F)^2
=(\hat{\bf{p}}-\mathbf{A}(\mathbf{r}))^2 +
\hbar B_z(\mathbf{r}) \sigma_z$ (see eq.(\ref{eq:second-H}))
in eq.(\ref{eq:w+n2}), 
we obtain
\begin{align}
 \mathbf{H}_{\rm K} = 
 {\cal H}_{\rm K} - 
 \gamma_n \left( \frac{{\cal H}_{\rm K}}{\gamma_0} \right)^2 
 + \gamma_n \frac{\ell^2}{\hbar} B_z(\mathbf{r}) \sigma_z.
 \label{eq:totH}
\end{align}

Having calculated the energy spectrum of ${\cal H}_{\rm K}$,
we can evaluate the energy shift due to the nnn interaction using
eq.(\ref{eq:totH}).
Because the edge state satisfies 
${\cal H}_{\rm K} \psi^{\rm K}_{0}(\mathbf{r}) = 0$ ($E=0$), 
the first-order energy shift is given by
\begin{align}
 \Delta E^{\rm nnn}_{\rm K} \equiv \gamma_n \frac{\ell^2}{\hbar}
 \int\!\!\int B_z(\mathbf{r}) \langle \sigma_z(\mathbf{r}) \rangle 
 d^2\mathbf{r},
 \label{eq:delE}
\end{align}
where 
$\langle \sigma_z(\mathbf{r}) \rangle \equiv (\psi^{\rm
K}_{0}(\mathbf{r}))^\dagger  \sigma_z \psi^{\rm K}_{0}(\mathbf{r})$
is the pseudo-spin density.
It follows from eqs.(\ref{eq:wfunc}), (\ref{eq:const-g(y)}),
(\ref{eq:g-ene}) and (\ref{eq:ans-gauge}) that
\begin{align}
 \int \langle \sigma_z(\mathbf{r}) \rangle dx =
 \left\{
 \begin{array}{@{\,}ll}
  + \frac{1}{\xi} e^{+\frac{2 y}{\xi}} & (y < 0), \\
  - \frac{1}{\xi} e^{-\frac{2 y}{\xi}} & (y > 0),
 \end{array} \right.
 \label{eq:p-spin-con}
\end{align}
which is nonzero only around the border, reflecting the 
pseudo-spin polarization.
Since $B_z(y)$ is confined near the border as shown in
Fig.~\ref{fig:graphene}(b), we assume
$B_z(y)=(\hbar/\ell) \delta (y-0_+)-(\hbar/\ell) \delta (y-0_-)$.
Putting eq.(\ref{eq:p-spin-con}) into eq.(\ref{eq:delE}), 
we obtain
\begin{align}
 \Delta E_{\rm K}^{\rm nnn} = - 2 \gamma_n \frac{\ell}{\xi},
\end{align}
which reproduces the previous result 
of the TB model for the edge state.~\cite{SMS}

An important feature of the nnn interaction is that 
the energy shift is always negative and therefore 
it always stabilizes the edge state.
The stability comes from 
the interaction between polarization of the pseudo-spin 
and the deformation-induced magnetic field.
Another important point is that 
the nnn interaction breaks the particle-hole symmetry.
As discussed previously, the edge state appears in pair 
($\psi^{\rm K}_E$ and $\sigma_z \psi^{\rm K}_E$)
at $E=0$ and their pseudo-spin densities are the same.
Thus, $\Delta E_{\rm K}^{\rm nnn}$ is the same for the two states,
and they remain degenerate at the energy 
$\Delta E^{\rm nnn}$ ($<0$) 
though the particle-hole symmetry is lost. 
It is clear in eq.(\ref{eq:totH}) that 
the particle-hole symmetry for the edge state 
is broken locally at the edge 
by the last term of the right-hand side.

Third, we consider an external magnetic field.
Since the sign in front of $\mathbf{A}^{\rm em}(\mathbf{r})$ is the same
for the K and K$'$ points as shown in eq.(\ref{eq:weyl-KK'}),
the energy shift by a magnetic field becomes
\begin{align}
 \Delta E^{\rm em} \equiv \gamma_n \frac{\ell^2}{\hbar}
  \int\!\!\int e B_z^{\rm em}(\mathbf{r})
  \langle \sigma_z(\mathbf{r}) \rangle d^2\mathbf{r}.
 \label{eq:delE-sub}
\end{align}
Thus, $B_z^{\rm em}(\mathbf{r})$ can 
shift the energy spectrum of the edge state.
When $B_z^{\rm em}(\mathbf{r})$ is uniform,
we estimate the energy shift as
$\Delta E^{\rm em} \approx \gamma_n ( \ell/\ell_B )^2$,
where $\ell_B$ is the magnetic length defined by 
$\ell_B = \sqrt{\hbar/eB^{\rm em}_z}$.
Here, we assume that eq.(\ref{eq:p-spin-con}) is still valid even in
the presence of a weak uniform magnetic field.
The numerical value of $\ell_B$ ($\ell$) 
is about $25/\sqrt{B^{\rm em}_z}$ nm (0.2 nm)
where the magnetic field is measured in the unit of Tesla.
For example, $B_z^{\rm em} \approx$ 10 Tesla gives a tiny shift
$\Delta E^{\rm em} \approx$ 0.2 meV. 
This value is one-order smaller than the Zeeman splitting.
Thus, compared with the deformation-induced magnetic field, 
external magnetic field has little effect on the edge state.

\section{Discussion and Summary}\label{sec:discuss}

By considering the edge state using the continuous model, 
we found that the deformation-induced gauge field 
($\mathbf{A}(\mathbf{r})$) 
and magnetic field ($B_z(\mathbf{r})$)
explain basic properties of the edge state.
It is summarized as follows:\\
(1) 
The  gauge field can generate the edge state in
energy spectrum, depending on the gauge field direction.
Let $\mathbf{e}_\parallel$ the unit vector along the border and 
$A_\parallel(\mathbf{r}) \equiv 
\mathbf{A}(\mathbf{r}) \cdot \mathbf{e}_\parallel$, 
localized states (the edge state) appears if 
the gauge field has a component parallel to the border: 
$A_\parallel(\mathbf{r}) \ne 0$. 
The edge states are pseudo-spin polarized.
\\
(2) The direction of the gauge field, namely, 
$A_\parallel(\mathbf{r}) > 0$ or $A_\parallel(\mathbf{r}) < 0$,
is vital for the edge states, 
as it determines the energy dispersion and wave vectors 
which allow the edge states. \\
(3) The flat energy band of edge states at zero energy 
results from a divergence of the gauge field; 
$A_\parallel(\mathbf{r}) \to \pm \infty$. \\
(4) The nnn interaction lowers the energy of the edge state 
below the Fermi energy. 
This is because the nnn induces a linear coupling 
between the pseudo-spin and $B_z(\mathbf{r})$.
\\
(5) $B_z(\mathbf{r})$ for the edge state corresponds to 
an enormous magnetic field $\sim 10^{5}$T at the border.
Thus a uniform external magnetic field has little effect on
the edge state, compared with the deformation.

Here, we discuss impurity effects on the edge states, as impurity
scattering may give rise to the Anderson localization for
low-dimensional systems.
However, it has been proposed that the Anderson localization does not
occur in graphene, since graphene has a 
remarkable property, leading to the absence of back-scattering.~\cite{ANS}
Because a rotation of the wave vector around the Fermi
point gives an additional phase shift of $\pi$ (Berry's phase),
a time-reversal pair of scattered waves cancel with each
other.~\cite{ANS}
However, in the presence of deformations such as a vacancy,
the wave function gets an extra phase shift through the AB effect of the
deformation-induced gauge field.
The extra phase makes an asymmetry between time-reversal pair of
scattering waves and thus the back-scattering which contributes to the
localization will be recovered.

Finally, we point out a possible experiment to confirm the results of
the present theory. 
A direct way of making localized states is by applying a stress locally
to an undeformed graphene, 
thereby inducing a deformation-induced magnetic field.
The present theory predicts that 
a formation of localized states depends sensitively on 
the direction of the stress, 
and such localized states can be probed by a peak in the local 
density of states in STS measurement.

In summary, 
we have formulated the edge state as a localized solution 
of the continuous model for a deformed graphene.
We found that the deformation-induced gauge field at the border 
is relevant to the presence of the edge state.
The gauge field can reproduce basic properties of the edge state;
the pseudo-spin polarization, the dispersion, and the flat energy band.
The gauge field is induced by local lattice deformation in general and
also can be simulated by an external magnetic field.
Compared with an external magnetic field, local lattice deformations
can generate a strong field of order of $10^5$ T.
The gauge field description was extended to include the SR 
and the nnn interaction. 
In terms of the gauge field,
we showed that the SR does not shift the energy of the edge state and 
the nnn stabilizes the edge state energy.

\section*{Acknowledgment}

K. S. acknowledges support form the 21st Century COE Program of the
International Center of Research and Education for Materials of Tohoku
University.
S. M. is supported by Grant-in-Aid (No.~16740167) from the Ministry of
Education, Culture, Sports, Science and Technology (MEXT), Japan.
R. S. acknowledges a Grant-in-Aid (No. 16076201) from MEXT.

\appendix
\section{Gauge Field for Weak Lattice Deformation}\label{app:gauge}

First, we derive the gauge field quoted in eq.(\ref{eq:gauge}).
It can be obtained by considering
the matrix element of the deformed Hamiltonian
between the Bloch wave functions.
The deformed Hamiltonian is defined by 
\begin{align}
 {\cal H}_{\rm deform} \equiv 
 \sum_{i \in {\rm A}} \sum_{a=1,2,3} \delta \gamma_a(\mathbf{r}_i) 
 (c_{i+a}^{\rm B})^\dagger c_i^{\rm A} + {\rm h.c.},
\end{align}
where $c_i$ and $c_i^\dagger$ are the canonical 
annihilation-creation operators of the electrons at site $i$.
The off-diagonal matrix element of ${\cal H}_{\rm deform}$ 
is given by
\begin{align}
 & \langle \Psi_{\rm A}^{\mathbf{k} + \delta \mathbf{k}} |
 {\cal H}_{\rm deform}|
 \Psi_{\rm B}^{\mathbf{k}} \rangle \nn \\
 & = \frac{1}{N_u} \sum_{i \in {\rm A}} \sum_{a=1,2,3} 
 \delta \gamma_a(\mathbf{r}_i) f_a(\mathbf{k}) e^{-i\delta \mathbf{k}
 \cdot \mathbf{r}_i},
 \label{app:matele}
\end{align}
where $f_a(\mathbf{k})\equiv e^{i\mathbf{k} \cdot \mathbf{R}_a}$ and
$\mathbf{R}_a$ ($a=1,2,3$) are vectors pointing to the nearest-neighbor
B sites from an A site.
Here, $|\Psi_s^{\mathbf{k}} \rangle$ is the Bloch wave function
defined as 
\begin{align}
 & |\Psi_s^{\mathbf{k}} \rangle = \frac{1}{\sqrt{N_u}}
 \sum_{i \in s} 
 e^{i \mathbf{k} \cdot \mathbf{r}_i} c_i^\dagger |0 \rangle, \ \ 
 (s = {\rm A},{\rm B})
\end{align}
where $N_u$ denotes the number of graphene unit cells.

By expanding $f_a(\mathbf{k})$ in eq.(\ref{app:matele})
around $\mathbf{k}_F$ (wave vector of the K point), we
obtain $f_a(\mathbf{k}) = f_a(\mathbf{k}_F) +
if_a(\mathbf{k}_F)(\mathbf{k}-\mathbf{k}_F) \cdot 
\mathbf{R}_a + \cdots$.
The second term and the higher order corrections can be ignored if
we consider the low-energy properties near the Fermi points 
of a deformed graphene 
satisfying $|\delta \gamma_a(\mathbf{r})| \ll \gamma_0$.
Then the dominant contribution is given by
\begin{align}
 \frac{1}{N_u} \sum_{i \in {\rm A}} \sum_{a=1,2,3} 
 \delta \gamma_a(\mathbf{r}_i) f_a(\mathbf{k}_F) e^{-i\delta \mathbf{k}
 \cdot \mathbf{r}_i}.
 \label{app:g-basic}
\end{align}
The K point satisfies
$\mathbf{k}_F \cdot \mathbf{a}_1 = -4\pi/3$ and 
$\mathbf{k}_F \cdot (\mathbf{a}_2-\mathbf{a}_1/2) = 0$ (mod $2\pi$), 
and therefore we obtain $f_1(\mathbf{k}_F) = 1$, 
$f_2(\mathbf{k}_F)=e^{+i\frac{2\pi}{3}}$ and 
$f_3(\mathbf{k}_F)=e^{-i\frac{2\pi}{3}}$.
Substituting these into eq.(\ref{app:g-basic}),
we see by the definition of 
$\mathbf{A}(\mathbf{r})$ (eq.(\ref{eq:gauge}))
\begin{align}
 & \langle \Psi_{\rm A}^{\mathbf{k} + \delta \mathbf{k}} |
 {\cal H}_{\rm deform}|
 \Psi_{\rm B}^{\mathbf{k}} \rangle \nn \\
 & \approx \frac{v_F}{N_{u}} 
 \sum_{i \in {\rm A}}
 \left\{
 A_x(\mathbf{r}_i) +i A_y(\mathbf{r}_i) \right\}
 e^{-i\delta \mathbf{k} \cdot \mathbf{r}_i}.
 \label{app:off-ele}
\end{align}
By the same calculation for 
$\langle \Psi_{\rm B}^{\mathbf{k} + \delta \mathbf{k}} | 
{\cal H}_{\rm deform}| \Psi_{\rm A}^{\mathbf{k}} \rangle$, 
we obtain the matrix for ${\cal H}_{\rm deform}$ as 
$-v_F \mathbf{\sigma}' \cdot \mathbf{A}(\mathbf{r})$.
Similarly, 
by expanding the nearest-neighbor Hamiltonian 
for an undeformed graphene around the K point, 
we see $v_F \mathbf{\sigma}' \cdot \hat{\mathbf{p}}$.
Thus, we obtain
$v_F \mathbf{\sigma}' \cdot (\hat{\mathbf{p}}-\mathbf{A})$, 
which is eq.(\ref{eq:weyl}).

\section{Derivation of eq.(\ref{eq:lat-res-fin})}\label{app:valid}

\begin{figure}[htbp]
 \begin{center}
  \includegraphics[scale=0.45]{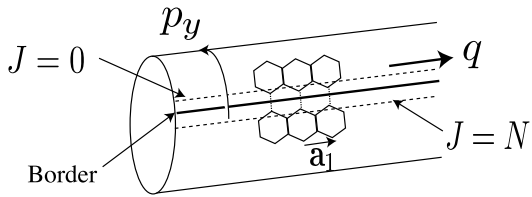}
 \end{center}
 \caption{Geometry of graphene with the zigzag border.
 When $c=1$, the system becomes a graphene with zigzag edges since the
 hopping integral across the border is $-(1-c)\gamma_0$.
 }
 \label{fig:band}
\end{figure}

Next, we consider a general border using the parameter $c$
to obtain eq.(\ref{eq:lat-res-fin}).
We write the Hamiltonian ${\cal H}_0$ 
for an undeformed periodic graphene 
shown in Fig.~\ref{fig:band} as,
\begin{align}
 {\cal H}_0 \equiv - \gamma_0 
 \sum_{a=1,2,3} \sum_{i \in {\rm A}} 
 (c_{i+a}^{\rm B})^\dagger c^{\rm A}_i + {\rm h.c.}
 \label{eq:H0}
\end{align}
The Hamiltonian for graphene with
zigzag edge is denoted by ${\cal H}_{\rm zigzag}$,
and we define ${\cal H}_{\rm border}$ by
${\cal H}_0 -{\cal H}_{\rm zigzag}$.
${\cal H}_{\rm border}$ describes the hopping across the border.
From the definition, ${\cal H}_{\rm border}$ is given by
\begin{align}
 {\cal H}_{\rm border} = -\gamma_0 \sum_{I}
 (c^{\rm A}_{I,J=N})^\dagger c^{\rm B}_{I,J=0} + {\rm h.c.},
 \label{eq:Hcut}
\end{align}
where $c_{I,J}$ and $c_{I,J}^\dagger$ are 
annihilation-creation operators of $\pi$-electron at the edge site, 
$(I,J)$ ($J=0$ or $N$).
In the following, we consider 
${\cal H}(c) \equiv {\cal H}_0 - c {\cal H}_{\rm border}$ where
$|c| \ll 1$ corresponds to a weak deformation and $c \to 1$
leads to ${\cal H}_{\rm zigzag}$, i.e., zigzag edge formation.
In addition to the zigzag edge~\cite{Fujita} for $c=1$,
${\cal H}(c)$ describes the Klein's edge~\cite{Klein} 
when $c\to -\infty$.
When $c\to -\infty$, an electron cannot come to the border sites,
and the system reduces to a graphite ribbon with the Klein's edges
which are made by removing the border sites from the zigzag edges.

Before we examine localized state of ${\cal H}(c)$ analytically,
let us show the numerical result of the energy band structure 
around the K point including both localized state and extended state.
In Figs.~\ref{fig:cone}(c),~\ref{fig:cone}(d) and~\ref{fig:cone}(e), 
we plot the energy spectrum of
${\cal H}(c)$ for $N=10$ with $c=0$, $1/2$, and $1$.
Because ${\cal H}_{\rm border}$ does not break the translational symmetry
along the border, the wave vector, 
$q \equiv \mathbf{k} \cdot \mathbf{a}_1$, remains to be a good quantum
number.
We consider the spectrum near $q = 2\pi/3$ (the K point). 
In the energy spectrum of ${\cal H}(0)$ (Fig.~\ref{fig:cone}(c)), the
valence and conduction bands touch at $q = 2\pi/3$.
Only extended states exist in the system and they are doubly degenerate,
which are supported by the (inversion) symmetry under
$p_y \leftrightarrow -p_y$ 
where $p_y$ denotes the momentum perpendicular to the border.
These degeneracies disappear as a result of the deformation, and 
the spacing of energy bands for ${\cal H}(1)$ 
becomes about one half of that for ${\cal H}(0)$. 
Apart from the change of energy spacing, we observe no significant
changes in the band structure of the extended states.

On the other hand, the localized states have several notable
characteristics.
First, localized state appears for $2\pi/3 < q \le \pi$ 
as we change $c$ from 0 to 1, i.e., 
there is an asymmetry about the K point in the momentum axis.
Second, the two localized states gradually 
depart from the valence and conduction bands 
respectively as $c$ is increased (Fig.~\ref{fig:cone}(d)).
Third, those localized states merge when $c=1$ 
to become the flat bands with zero energy (Fig.~\ref{fig:cone}(e)).
In the following, 
we will analyze these features by analytically calculating 
the energy eigenvalue and localization length of localized
state.

An energy eigenstate can be represented as
\begin{align}
 \psi_{E}(I,J) = N' \exp(iq I) 
 \begin{pmatrix}
  \phi_{\rm A}(J) \cr \phi_{\rm B}(J)
 \end{pmatrix}.
\end{align}
The energy eigenequation ${\cal H}(c)\psi_{E}=E\psi_{E}$ 
is written as 
\begin{align}
 \begin{split}
  & \epsilon \phi_{\rm A}(J) = - \phi_{\rm B}(J+1) - Q \phi_{\rm B}(J), \\
  & \epsilon \phi_{\rm B}(J+1) = - \phi_{\rm A}(J) -Q \phi_{\rm A}(J+1),
 \end{split}
 \label{app:ene-ign}
\end{align}
where $Q \equiv 2 \cos (q/2)$ and 
$\epsilon$ ($\equiv E/\gamma_0$) is the energy eigenvalue 
normalized by the hopping integral.
The eigenstate is expressed as
\begin{align}
 &
 \begin{pmatrix}
  \phi_{\rm A}(J) \cr \phi_{\rm B}(J)
 \end{pmatrix}
 = S_J
 \begin{pmatrix} 
  Q +\frac{S_{J+1}}{S_J} & \epsilon \cr
  -\epsilon & -\left( Q +\frac{S_{J-1}}{S_J} \right)
 \end{pmatrix}
 \begin{pmatrix}
  \phi_{\rm A}(0) \cr \phi_{\rm B}(0)
 \end{pmatrix}, \nonumber \\
 & \ \ \ \ 
 (J=0,\ldots,N) \label{eq:wf}
\end{align}
where $S_J \equiv \sin (J \phi)/\sin \phi$.
Here $\phi$ is a wavenumber in the direction of $J$ 
and satisfies
\begin{align}
 \epsilon^2 = Q^2 + 2Q \cos \phi + 1.
 \label{eq:energy-tb}
\end{align}
The derivation of eq.~(\ref{eq:wf}) is as follows:
first, it is clear from eq.(\ref{app:ene-ign}) 
that eq.(\ref{eq:wf}) holds for $J=0$ and  $J=1$.
Second, let us rewrite eq.(\ref{app:ene-ign}) using
eq.(\ref{eq:energy-tb}) as 
\begin{align}
 \begin{pmatrix}
  \phi_{\rm A}(J+1) \cr \phi_{\rm B}(J+1)
 \end{pmatrix} = 
 \begin{pmatrix}
  Q+2\cos \phi & \epsilon \cr
  -\epsilon & -Q
 \end{pmatrix}
 \begin{pmatrix}
  \phi_{\rm A}(J) \cr \phi_{\rm B}(J)
 \end{pmatrix}.
 \label{eq:mat-sec}
\end{align}
Equation~(\ref{eq:wf}) can then be proved by induction.

The wave function is given by three parameters $c$, $q$ and $\phi$.
The parameter $\phi$ is determined by the boundary condition: 
the energy eigenequation at border sites, 
$J=0$ (B sites) and $J=N$ (A sites), 
\begin{align}
 \begin{split}
  & \epsilon \phi_{\rm B}(0) 
  = -(1-c) \phi_{\rm A}(N) - Q \phi_{\rm A}(0), \\ 
  & \epsilon \phi_{\rm A}(N)
  = -(1-c) \phi_{\rm B}(0) - Q \phi_{\rm B}(N).
 \end{split}
 \label{eq:dy-b-pre}
\end{align}
Equation~(\ref{eq:dy-b-pre}) can be rewritten as
\begin{align}
 \begin{pmatrix}
  \phi_{\rm A}(N) \cr \phi_{\rm B}(N)
 \end{pmatrix}
 = \frac{1}{1-c}
 \begin{pmatrix}
  -Q & -\epsilon \cr
  \epsilon & \frac{\epsilon^2 -(1-c)^2}{Q} 
 \end{pmatrix}
 \begin{pmatrix}
  \phi_{\rm A}(0) \cr \phi_{\rm B}(0)
 \end{pmatrix}.
 \label{eq:dy-b}
\end{align}
From eqs.(\ref{eq:wf}) and (\ref{eq:dy-b}), we obtain two equations
relating 
$(\phi_{\rm A}(0),\phi_{\rm B}(0))$ to $(\phi_{\rm A}(N),\phi_{\rm B}(N))$.
Then, by eliminating $\phi_{\rm A}(N)$ and $\phi_{\rm B}(N)$ in these
equations, we obtain the following constraint equation for $\phi$,
\begin{align}
 \det
 \begin{pmatrix}
  (Q+\frac{S_{N+1}}{S_N})+\frac{Q}{A_N} & \epsilon (1+\frac{1}{A_N})
  \cr
  -\epsilon (1+\frac{1}{A_N}) &
  -(Q+\frac{S_{N-1}}{S_N})-\frac{\Delta}{A_N} 
 \end{pmatrix}
 =0,
 \label{eq:cons}
\end{align}
where
\begin{align}
 & A_N \equiv (1-c)S_N, 
 \ \ \Delta \equiv \frac{\epsilon^2-(1-c)^2}{Q}.
 \label{eq:subvaris}
\end{align}
Equation~(\ref{eq:cons}) has
$2N$ solutions of $\phi$ for given $N$.
From the numerical calculation, we find that for 
$|Q| > 1$, namely, for $0 < q < 2\pi/3$ or $4\pi/3 < q < 2\pi$, 
the equation has $2N$ real value solutions of $\phi$,
and for $|Q| <1$ ($2\pi/3 < q < 4\pi/3$), 
it has $2(N-1)$ real value solutions (extended states) 
and 2 complex value solutions (localized states).
As for localized state, $\phi$ can be parameterized as 
$\phi = i \ell/\xi$ or $\phi = \pi + i \ell/\xi$ 
depending on $-1<Q<0$ and $0<Q<1$, respectively.
Here, $\xi$ is the localization length and thus 
when $\xi \to \infty$ the state becomes an extended state.
From eq.(\ref{eq:energy-tb}), we have $\epsilon^2 = (Q \pm 1)^2$ for
$\phi=0$ and $\pi$.
Thus, one can understand that there are two points in $(q,\phi)$
plane satisfying $\epsilon=0$, where a localized state turns into an
extended state.~\cite{SMSK}
One point is located at $(2\pi/3,\pi)$ ($Q=1$) which is the K
point and the other is located at $(4\pi/3,0)$ ($Q=-1$) which is the
K$'$ point.

When $N\gg 1$,
by solving eq.(\ref{eq:cons}) for localized state, 
we acquire 
\begin{align}
 \frac{\ell}{\xi} = \ln \left( 
 \frac{c(2-c)+\sqrt{c^2(2-c)^2+4Q^2(1-c)^2}}{2|Q|}
 \right)+\cdots,
 \label{eq:local-ana}
\end{align}
where $\cdots$ denotes correction of ${\cal O}(e^{-N \ell/\xi})$ and 
therefore it is negligible when $N \gg \xi/\ell$.
To obtain eq.(\ref{eq:local-ana}),
we neglected $A_N^{-2}$ in eq.(\ref{eq:cons}) since 
$A_N$ is large as ${\cal O}(e^{N \ell/\xi})$.
In this approximation, we get 
$Q(Q+e^{-\ell/\xi})+\Delta(Q+e^{\ell/\xi})=2\epsilon^2$ for $Q < 0$.
Using eqs.(\ref{eq:energy-tb}) and (\ref{eq:subvaris}), 
we rewrite this condition as
\begin{align}
 (e^{\ell/\xi})^2 - \frac{c(c-2)}{Q}e^{\ell/\xi} - (1-c)^2 = 0,
 \label{eq:loc-der-c}
\end{align}
which gives eq.(\ref{eq:local-ana}).
The energy eigenvalue is obtained by inserting
eq.(\ref{eq:local-ana}) into eq.(\ref{eq:energy-tb}).
It is to be noted that not only $c=0$ but also $c=2$ give no
localized state since eq.(\ref{eq:loc-der-c}) gives $\ell/\xi=0$
in these cases.
For $c=2$, 
the hopping integral across the border is $+\gamma_{0}$, 
while other bonds are $-\gamma_{0}$.
By a gauge transformation, this difference in sign of the hopping 
integral on the border can be absorbed into an AB phase $\pi$ 
around the nanotube (to the $y$-direction in Fig.~\ref{fig:band}).
Thus the system is equivalent to a 
SWNT without border, with a $\pi$ flux 
(a half of the flux quantum) parallel to the
NT axis. Because it has no border, it has no edge state.

Next, to understand $E$ and $\xi$ for localized state near the K point,
we define the wave vector $k_x$ measured from K point as 
$q=2\pi/3+|\mathbf{a}_1|k_x$. 
Because $Q=2\cos(q/2)$, we have 
$Q = 1-\ell k_x + (\ell k_x)^2/6 + {\cal O}((\ell k_x)^3)$.
Then, from eq.(\ref{eq:energy-tb}) with $\phi = \pi+i \ell/\xi$, 
we obtain 
\begin{align}
 \left( \frac{E}{v_F} \right)^2 = 
 p_x^2 - \left( \frac{\hbar}{\xi} \right)^2 + \cdots,
 \label{eq:ene-dis-lattice-appr}
\end{align}
where $\cdots$ indicates higher order corrections such as 
$\ell p_x^3/\hbar$.
The first two dominant terms in the right-hand side 
reproduce eq.(\ref{eq:energydispersion}).
By expanding eq.(\ref{eq:local-ana}) in terms of $p_x (=\hbar k_x)$, 
we obtain eq.(\ref{eq:lat-res-fin}).

\begin{figure}[htbp]
 \begin{center}
  \includegraphics[scale=0.45]{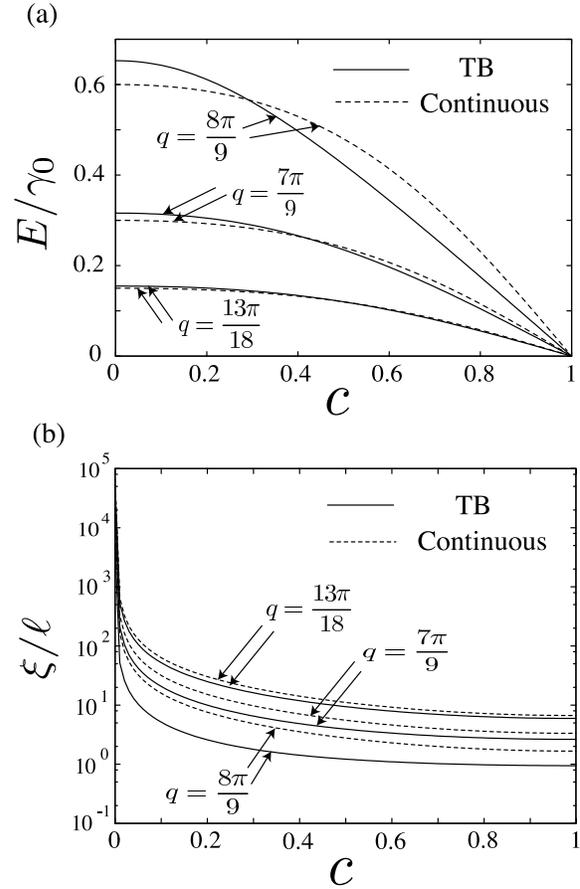}
 \end{center}
 \caption{(a) Energy eigenvalue $E/\gamma_0$
 and (b) localization length $\xi/\ell$ as a function of $c$.
 The solid and dashed curves represent the TB model 
 and the continuous model, respectively.
 Three wave vectors $q=13\pi/18$, $7\pi/9$ and
 $8\pi/9$ are chosen as examples.
 When $c=1$, each curve in (a) converges to zero, 
 which agrees with the flat band of edge states, and 
 each solid curve in (b) reproduces 
 previous results of the localization length.~\cite{Fujita,SMSK}
 }
 \label{fig:1-hikaku}
\end{figure}

Finally, we check whether $E$ and $\xi$ of the continuous model
reproduce the results of the TB model for localized states
satisfying $k_x \ell \approx 1$.
First, using eqs.(\ref{eq:energy-tb}) and (\ref{eq:local-ana}), 
we plot $E$ and $\xi$ for $q=13\pi/18$, $7\pi/9$ and $8\pi/9$ 
as solid curves 
in Figs.~\ref{fig:1-hikaku}(a) and~\ref{fig:1-hikaku}(b), respectively.
In Fig.~\ref{fig:1-hikaku}(a), 
$E$ decreases slowly at small $c$.
However, every curve decreases almost linearly as a function of $c$ 
for large $c$ and converges to zero.
In Fig.~\ref{fig:1-hikaku}(b) we see that 
$\xi$ decreases quickly with increasing $c$ and converge
at $\xi/\ell= -1/\ln|Q|$ (eq.(\ref{eq:local-ana}) with $c=1$).
The convergence agrees with the previously published
literatures.~\cite{Fujita,SMSK}
The dashed lines 
in Figs.~\ref{fig:1-hikaku}(a) and~\ref{fig:1-hikaku}(b)
are $E$ and $\xi$ derived from the continuous model.
The wave vectors $q=13\pi/18$, $7\pi/9$ and $8\pi/9$, which we used 
in the plot for the TB model, 
correspond to $|\mathbf{a}_1|k_x = \pi/18$,
$\pi/9$ and $2\pi/9$, in the continuous model.
Using eqs.(\ref{eq:ene}) and (\ref{eq:ans-gauge}), 
we plot $E/\gamma_0$ ($=\epsilon$) as dashed curves 
in Fig.~\ref{fig:1-hikaku}(a). 
One can see that the solid and dashed
curves are almost identical for $q=13\pi/18$.
However, their difference is visible for $q=7\pi/9$ and not negligibly
small for $q=8\pi/9$.
Using eqs.(\ref{eq:xi}) and (\ref{eq:ans-gauge}),
we plot $\xi/\ell$ as the dashed curves 
in Fig.~\ref{fig:1-hikaku}(b).
The continuous and the TB models agree with each other 
with $c$ near zero.  
For general $c$ and $q$, 
while the trend of $\xi$ agrees for the two models, 
their difference becomes large as
the state is located apart from the K point.
The discrepancy between the TB and continuous models comes from 
the assumption of the linear $k$ energy dispersion relation 
of the Weyl equation.
The difference increases as the state is located apart from the
Fermi point, which represents the deviation of the energy band from the 
linear energy dispersion.
The energy dispersion relation may be improved to reproduce the TB
model by adding some higher order terms to the Weyl equation.

\end{document}